\documentclass[12pt,a4paper]{article}
\usepackage{a4}

\ifx\pdfoutput\undefined
\usepackage{hyperref}
\usepackage{graphicx}
\else
\usepackage[pdftex]{hyperref}
\usepackage[pdftex]{graphicx}
\fi
\graphicspath{{./figures/}}

\newcommand{\ep}{\epsilon}
\newcommand{\als}{\alpha_s}
\let\*=\,
\newcommand{\nl}{n_l}
\newcommand{\nh}{n_h}
\newcommand{\nf}{n_f}
\newcommand{\vep}{\varepsilon}
\newcommand{\lp}{\ell}
\newcommand{\Order}{\mathcal{O}}
\newcommand{\z}[1]{\zeta_{#1}}
\newcommand{\pl}[1]{a_{#1}}
\newcommand{\dl}[1]{{d^d\lp_{#1}\over(2\*\pi)^d}}
\newcommand{\Log}[2]{\log^{#2}({#1})}
\newcommand{\lmusdms}{l_{\mu}}
\begin{document}

\begin{titlepage}
\noindent
\hfill SFB/CPP--04--68\\
\mbox{}
\hfill TTP04--25\\
\mbox{}

\vspace{0.5cm}
\begin{center}
  \begin{Large}
    \begin{bf}
      Heavy-quark vacuum polarization: first two moments of 
      the  {\boldmath{$\Order(\als^3 n_f^2)$}} contribution 
    \end{bf}
  \end{Large}
  \vspace{0.8cm}

  \begin{large}
    K.G. Chetyrkin$\rm \, ^{a \,}$\footnote{On leave from Institute for
      Nuclear Research of the Russian Academy of Sciences, Moscow, 117312,
      Russia.}, J.H.~K\"uhn$\rm \, ^{a \,}$, P.~Mastrolia$\rm \, ^{b \,}$
    {\normalsize and } C.~Sturm$\rm \, ^{a \,}$
  \end{large}
  \vskip .7cm

	 {\small {\em 
	     $\rm ^a$  Institut f\"ur Theoretische Teilchenphysik,
	     Universit\"at Karlsruhe,
	     D-76128 Karlsruhe, Germany}}
	 \vskip .3cm
	{\small {\em 
	    $\rm ^b$ Department of Physics and Astronomy, UCLA,
	    Los Angeles, CA 90095-1547}}
	
	\vspace{0.8cm}
{\bf Abstract}
\end{center}
\begin{quotation}
  \noindent
  The vacuum polarization due to a virtual heavy quark pair and
  specifically the coefficients of its Taylor expansion in the external
  momentum are closely related to moments of the cross section for
  quark-antiquark pair production in electron-positron
  annihilation. Relating measurement and theoretically calculated Taylor
  coefficients, an accurate value for charm- and bottom-quark mass can be
  derived, once corrections from perturbative QCD are sufficiently well
  under control. Up to three-loop order these have been evaluated
  previously. We now present a subset of four-loop contributions to the
  lowest two moments, namely those from diagrams which involve two internal
  loops from massive and massless fermions coupled to virtual gluons,
  hence of order $\als^3 n_f^2$. The calculation demonstrates the
  applicability of Laporta's algorithm to four-loop vacuum diagrams with
  both massive and massless propagators and should be considered a
  first step towards the full evaluation of the order $\als^3$
  contribution.
\end{quotation}
\end{titlepage}
\section{Introduction}
The correlator of two currents is central for many theoretical and
phenomenological investigations in Quantum Field Theory (QFT) (for a
detailed review see e.g.~\cite{ChKK:Report:1996,Steinhauser:2002rq}).
Important physical observables like the cross section of
electron-positron annihilation into hadrons and the decay rate of the
$Z$-boson are related to the vector and axial-vector current
correlators.  Total decay rates of CP even or CP odd Higgs bosons can be
obtained by considering the scalar and pseudo-scalar current densities,
respectively. Some of these studies require to calculate the correlator
for arbitrary momentum $q^2$. For many applications, however, the
knowledge of a few derivatives at $q^2=0$ is sufficient.

Two-point correlators have been studied in great detail in the framework
of perturbative QFT. Indeed, due to the simple kinematics (only one
external momentum) even multi-loop calculations can be performed
analytically. The results for all physically interesting diagonal and
non-diagonal correlators (vector, axial-vector, scalar and
pseudo-scalar) are available up to order $\als^2$, taking into account
the {\em full} quark mass and momentum dependence
\cite{Chetyrkin:1996cf,Chetyrkin:1998ix,Chetyrkin:1997mb}.

The determination of the heavy quark masses with the help of QCD sum
rules requires the detailed knowledge of the heavy quark correlators.
In fact, as noted in ref.~\cite{Kuhn:2001dm}, the precise determination
of the charm- and bottom-quark masses would be further improved by
including {\em four-loop corrections}, hence $\Order(\als^3)$, to at
least the few lowest Taylor coefficients of the
polarization function.

Technically speaking, the moments i.e. certain weighted integrals of the
cross section for electron-positron annihilation into heavy quarks, can
be expressed through massive tadpoles or vacuum diagrams (diagrams
without dependence on the external momentum).  The evaluation of these
``massive tadpoles'' in three-loop approximation has been pioneered in
\cite{Broadhurst:1991fi} and automated and applied to a large class of
problems in \cite{Steinhauser:2000ry}. However, in spite of huge progress
in calculational techniques during recent years the problem of
analytical calculation of massive tadpoles at the four-loop level has
not yet been mastered.\footnote{The numerical evaluation is certainly
possible for individual contributions; however, one could hardly imagine
a direct numerical evaluation of hundreds of thousands of separate terms
which appear after performing necessary expansions and traces at the
four-loop level.}

Similar to the three-loop case, the analytical evaluation of four-loop
tadpole integrals is based on the Integration-by-parts (IBP) method
\cite{Chetyrkin:1981qh}. In contrast to the three-loop case the manual
construction of algorithms to reduce arbitrary diagrams to a few master
integrals is replaced by a mechanical solution of a host of again
mechanically generated IBP equations \cite{Laporta:2001dd}.

Unfortunately, the price for this automatization is an enormous demand
on computational power. A system of more than ten million linear
equations has to be generated and solved. In the present publication we
present only a partial result. We restrict ourselves to the first two
non-vanishing moments and consider only four-loop diagrams with the
maximal number (two) of closed fermion loops inside.  This corresponds
to terms proportional to $\nl^2$, $\nh^2$ and $\nl\*\nh$, where $\nl$
denotes the number of light quarks, considered as massless and $\nh$ the
number of massive ones.  This leads to a system of about one million
equations.

In general, the tadpole diagrams encountered during our calculation
contain both massive and massless lines. As is well-known, the
computation of the four-loop $\beta$-functions requires the
consideration of four-loop tadpoles only composed of {\em completely
massive} propagators.  Calculations for this particular case have been
performed in \cite{vanRitbergen:1997va,Czakon:2004bu}.

The outline of this paper is as follows. In section
\ref{GeneralNotations} we briefly introduce the notation and discuss
generalities. In section \ref{Calculations} we discuss the reduction to
master integrals, describe the solution of the linear system of
equations and give the result for the $\Order(\als^3 n_f^2)$
contribution to the polarization function for the lowest two
moments. Our conclusions and a brief summary are given in section
\ref{DiscussConclude}.

\section{Notation and Generalities}

\label{GeneralNotations}
The vacuum polarization tensor $\Pi^{\mu\nu}(q^2)$ is defined as
\begin{equation}
  \Pi^{\mu\nu}(q^2)=i\*\int dx\,e^{iqx}\langle 0|Tj^\mu(x) j^\nu(0)|0 \rangle,
\end{equation}
where $q^{\mu}$ is the external momentum and $j^{\mu}$ is the
electromagnetic current.  The tensor $\Pi^{\mu\nu}(q^2)$ can be
expressed by a scalar function, the vacuum polarization function
$\Pi(q^2)$ through
\begin{equation}
  \label{vacpol}
  \Pi^{\mu\nu}(q^2)=\left(-q^2\*g^{\mu\nu}+q^{\mu}\*q^{\nu}\right)\*\Pi(q^2)+q^{\mu}\*q^{\nu}\*\Pi_L(q^2).
\end{equation}
The longitudinal part $\Pi_L(q^2)$ is equal zero due to the Ward
identity.\\ 
The confirmation of $\Pi^{\mu}_{\mbox{ }\mu}(q^2=0)=0$ and
$\Pi_{L}(q^2)=0$ will constitute an important check of our
calculation. The constant $\Pi(q^2=0)$ relates the QED coupling in the
on-shell scheme and the $\overline{\mbox{MS}}$-scheme.  The first and
higher derivatives of $\Pi(q^2)$ at $q^2=0$ contain important scheme
independent information and will sometimes be called the physical
moments.  The imaginary part of the polarization function is related to
the physical observable $R(s)$
\begin{equation}
  R(s)=12\*\pi\*\mbox{Im}\,\Pi(q^2=s+i\*\vep).
\end{equation}
and properly weighted integrals of $R(s)$ obviously coincide with the
Taylor coefficients
\begin{equation}
{1\over n!}\*\left({d\over dq^2}\right)^n\left.\Pi(q^2)\right|_{q^2=0}=
{1\over 12\*\pi^2}\int_{0}^{\infty} ds\;{ R(s)\over s^{n+1}} 
\end{equation}
This justifies to study the Taylor expansion around $q^2=0$. In this
case the coefficients are given by massive tadpole integrals  
\begin{equation}
  \Pi(q^2) = {N_c\over16\*\pi^2}\*\sum_{n\ge0} C_n\*z^n
\end{equation}
with the dimensionless variable $z=q^2/(4\*m^2)$, where $m$ is the mass
of the heavy quark and $N_c$ denotes the number of colors. It is
convenient to define the expansion of the  
coefficients $C_n$ of the polarization function in the strong coupling
constant $\als$ as
\begin{equation}
C_n=C_n^{(0)}+a_{s}\*C_n^{(1)}+a_{s}^2\*C_n^{(2)}+a_{s}^3\*C_n^{(3)}+\dots
\end{equation}
with $a_{s}={\als\over\pi}$. 
Within this work we consider the $\nf^2$ contribution of $C_n^{(3)}$ and
define $C_n^{(3)}\big|_{\nf^2}=T^2\*C_F\*\hat{C}_n^{(3)}$, where
$T$ denotes the normalization factor of the fundamental-representation
generators $t^a$ defined by $\mbox{Tr}[t^a\*t^b]=T\*\delta^{ab}$ and
$C_F$ is the Casimir operator in the fundamental representation.
\section{Calculations and Results}
\label{Calculations}
\subsection*{Reduction to master integrals}

In the first step the reducible scalar products in the numerator of the
integrands have been removed in the sense that trivial tensor reduction
has been performed. All reducible scalar products have been expressed
in terms of their associated denominators. Through this the remaining
integrals can be mapped upon a set of 12 independent topologies. 

\begin{figure}[h]
\begin{center}
\hspace{-0.5cm}
\begin{minipage}[t]{2.5cm}
\includegraphics[height=2.5cm,bb=71 106 684 720]{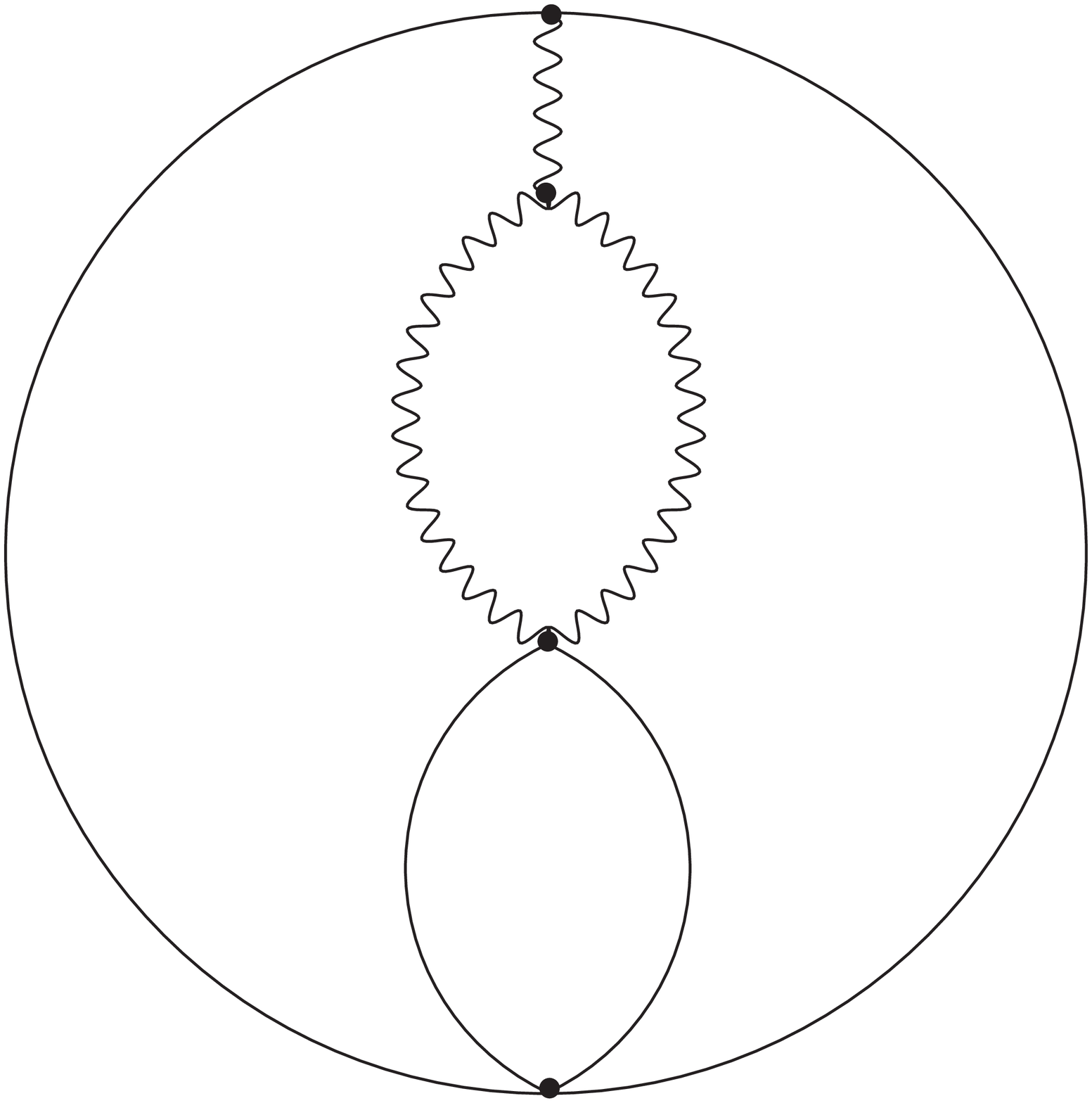}
\vspace{-0.5cm}
\begin{center}
$R_1$
\end{center}
\end{minipage}
\begin{minipage}[t]{2.5cm}
\includegraphics[height=2.5cm,bb=71 106 684 720]{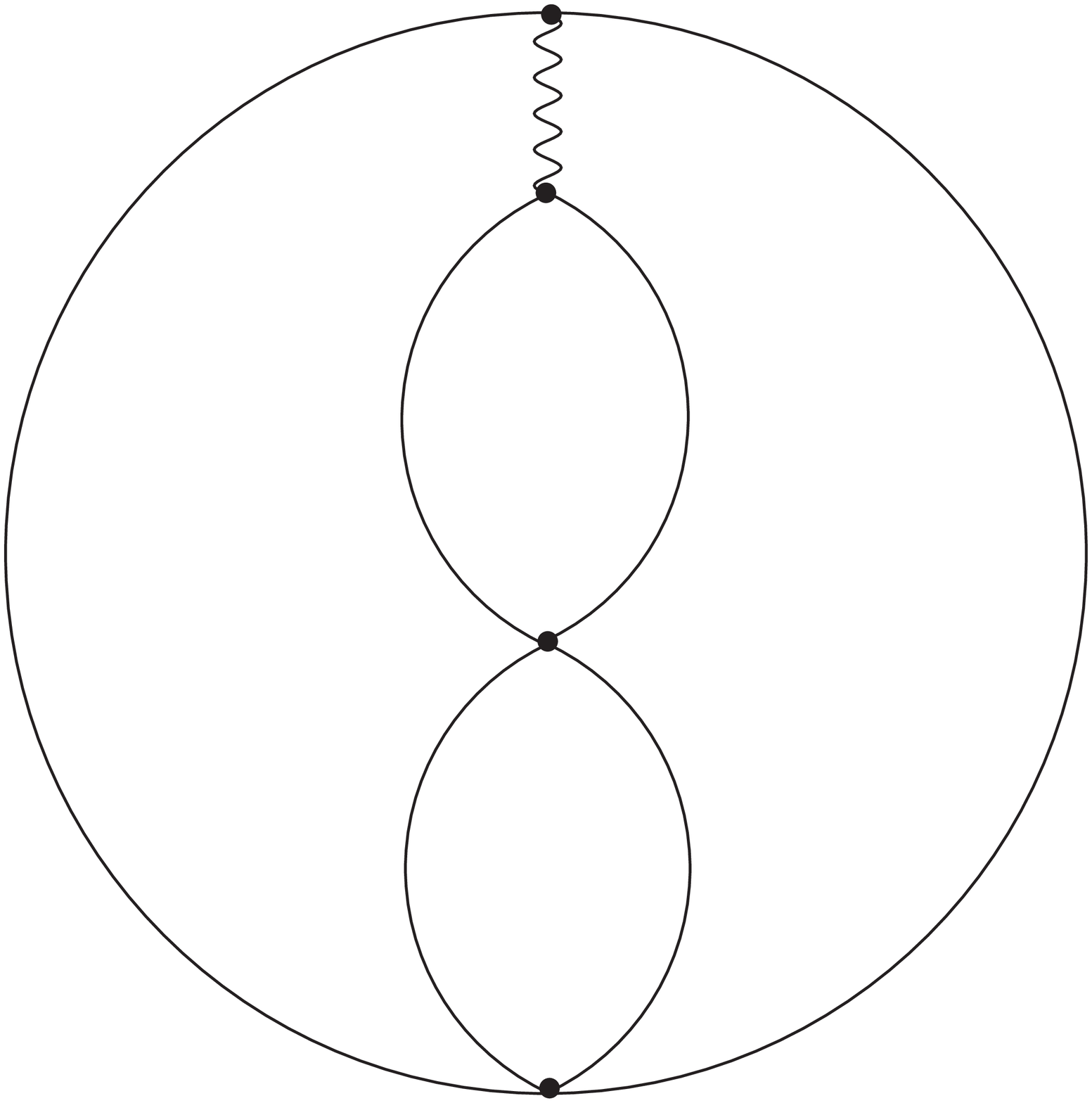}
\vspace{-0.5cm}
\begin{center}
$R_2$
\end{center}
\end{minipage}
\begin{minipage}[t]{2.5cm}
\includegraphics[height=2.5cm,bb=71 106 684 720]{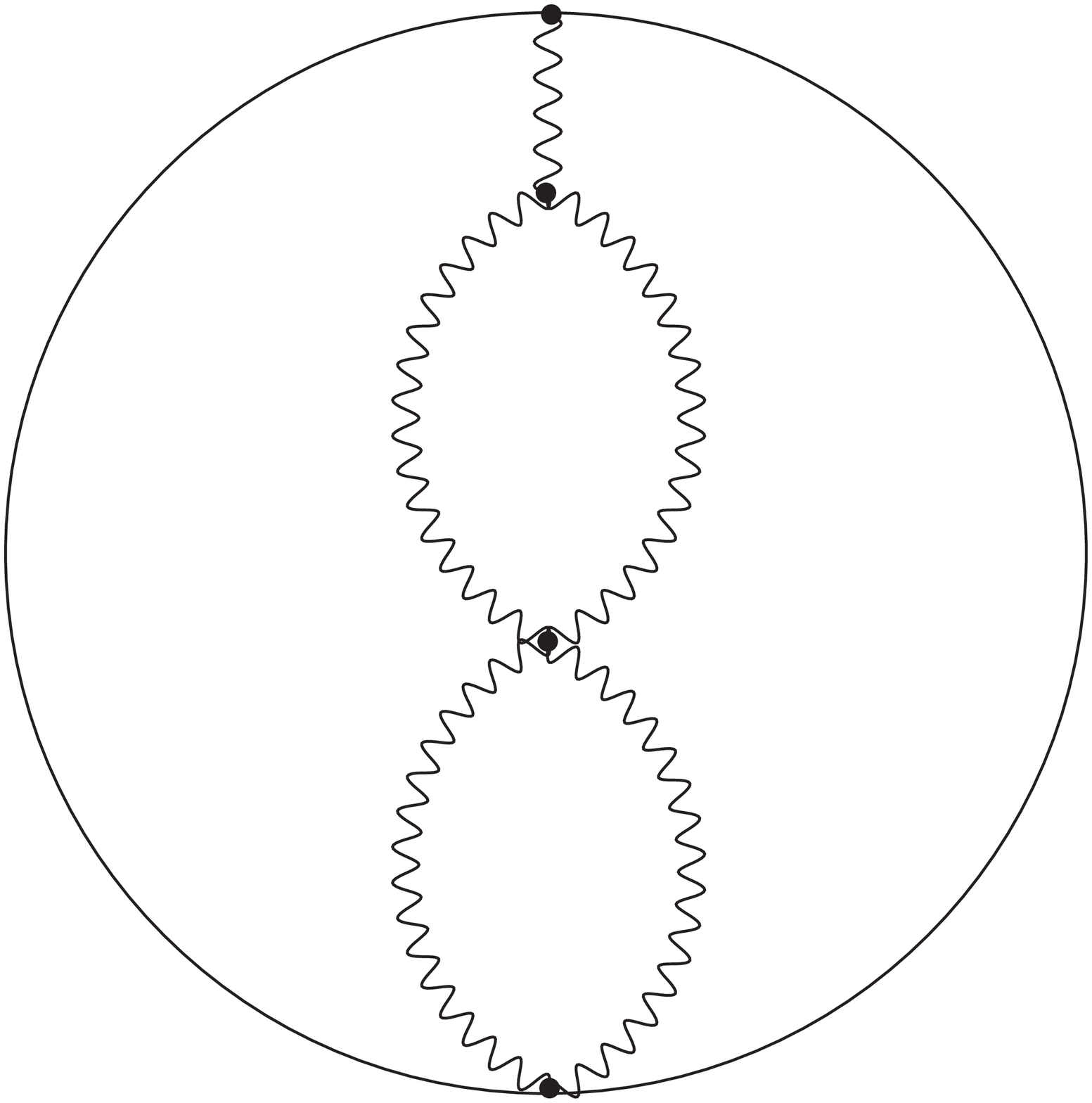}
\vspace{-0.5cm}
\begin{center}
$R_3$
\end{center}
\end{minipage}
\begin{minipage}[t]{2cm}
\includegraphics[height=2.5cm,bb=71 107 539 720]{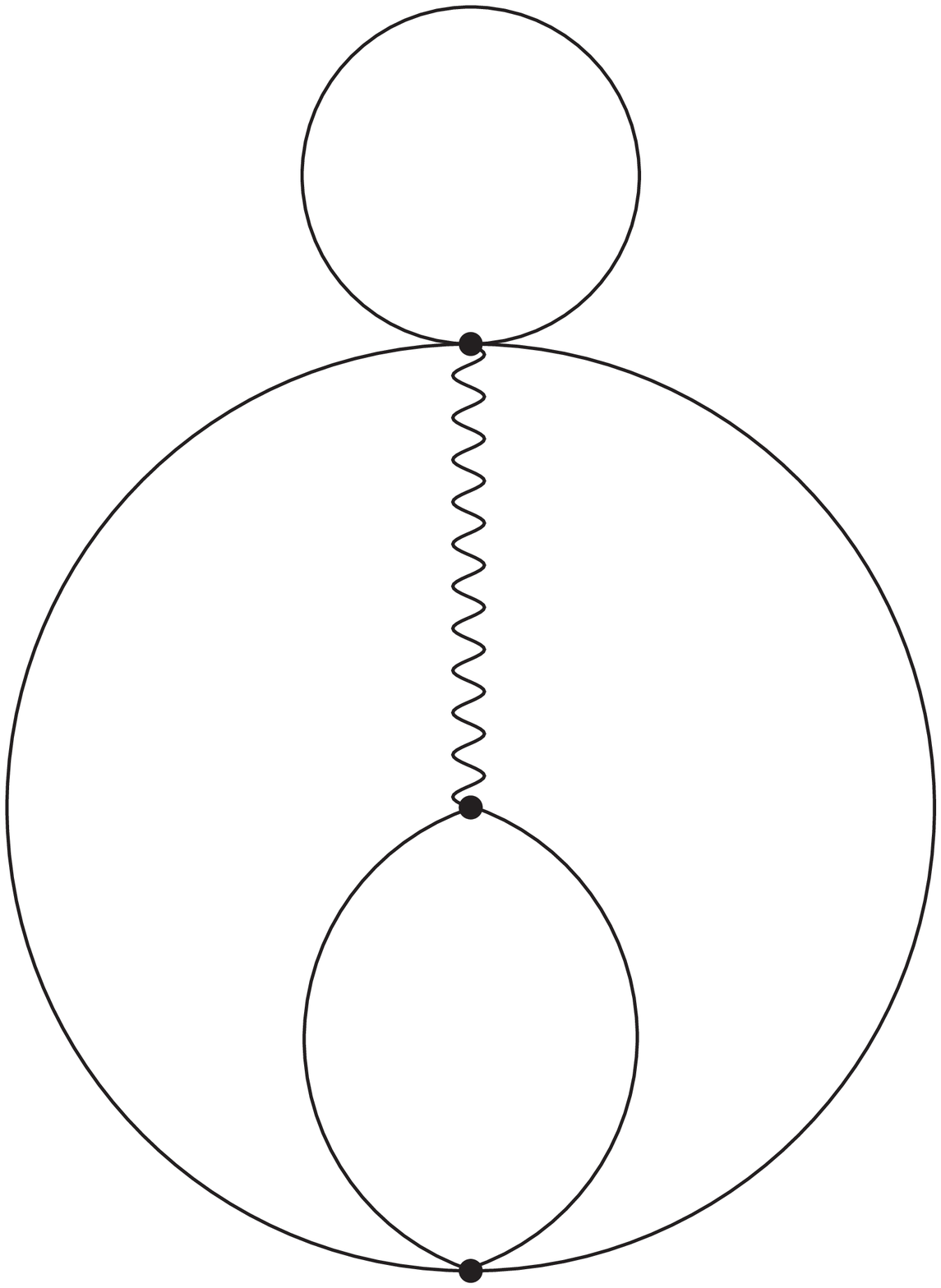}
\vspace{-0.5cm}
\begin{center}
$R_4$
\end{center}
\end{minipage}
\begin{minipage}[t]{2cm}
\includegraphics[height=2.5cm,bb=71 107 525 720]{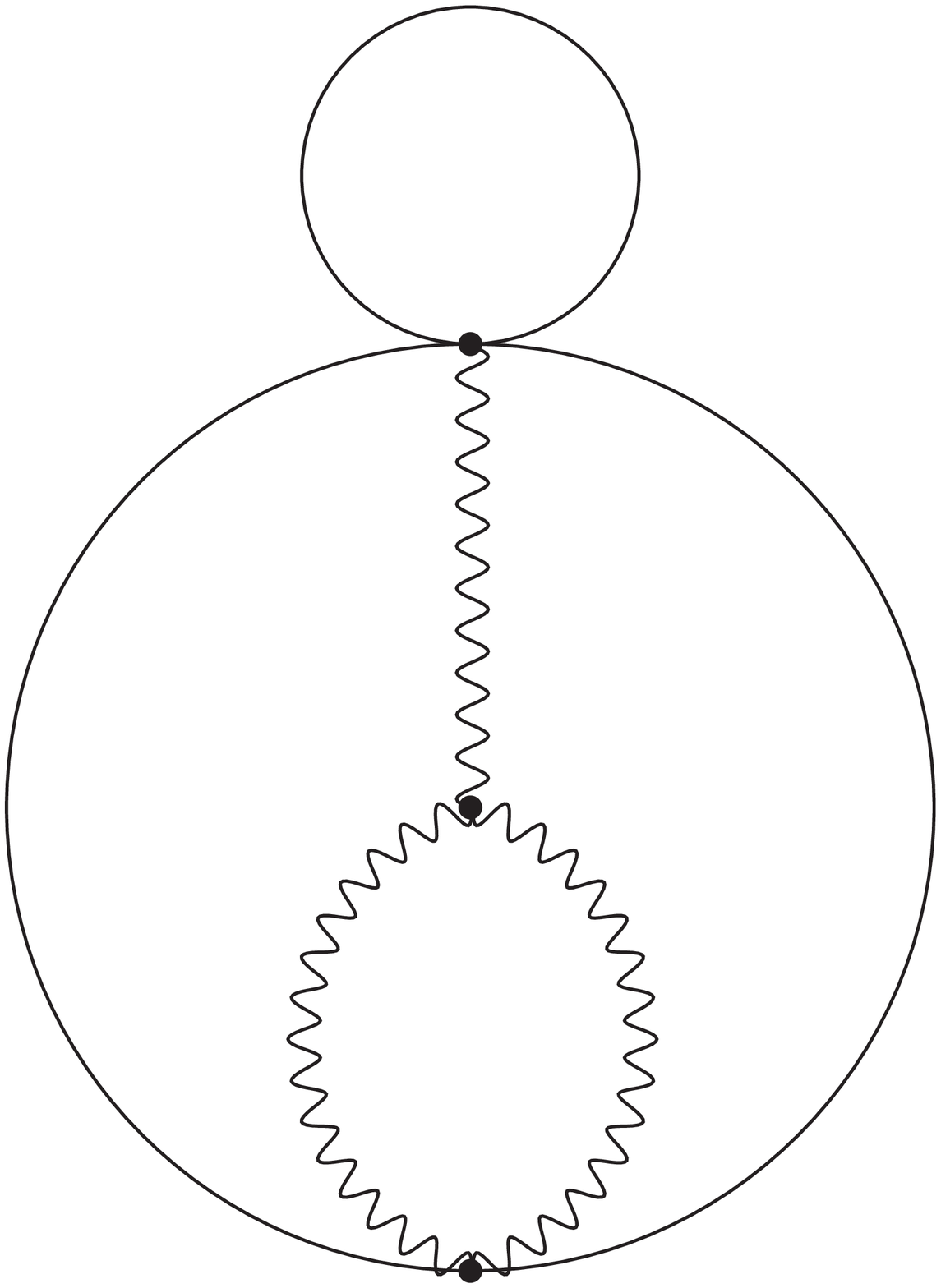}
\vspace{0.1cm}
\begin{center}
$R_5$
\end{center}
\end{minipage}
\begin{minipage}[t]{2cm}
\includegraphics[height=2.5cm,bb=119 115 477 721]{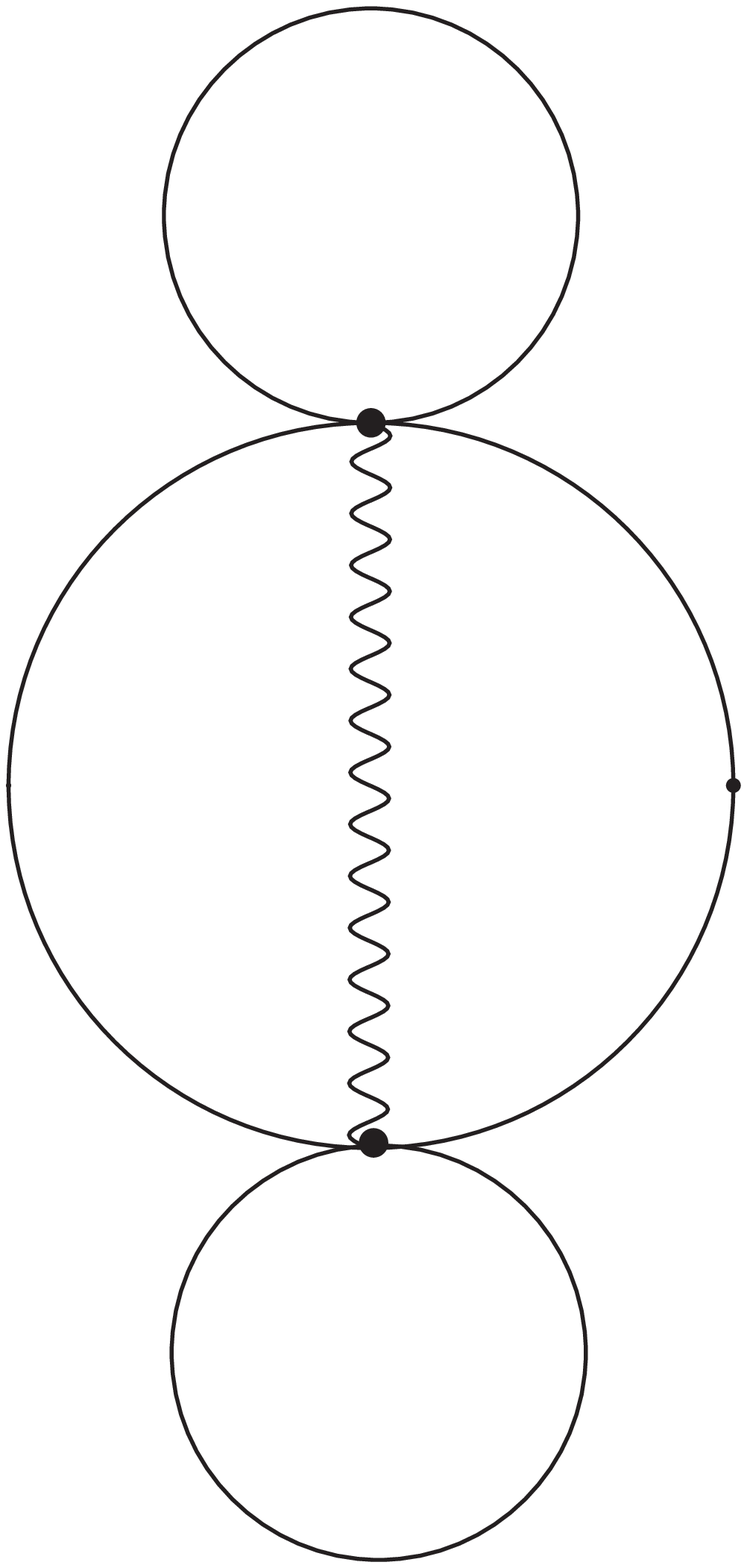}
\vspace{0.1cm}
\begin{center}
$R_6$
\end{center}
\end{minipage}
\end{center}
\begin{center}
\hspace{-0.5cm}
\begin{minipage}[t]{2.5cm}
\includegraphics[height=2.5cm,bb=71 107 525 720]{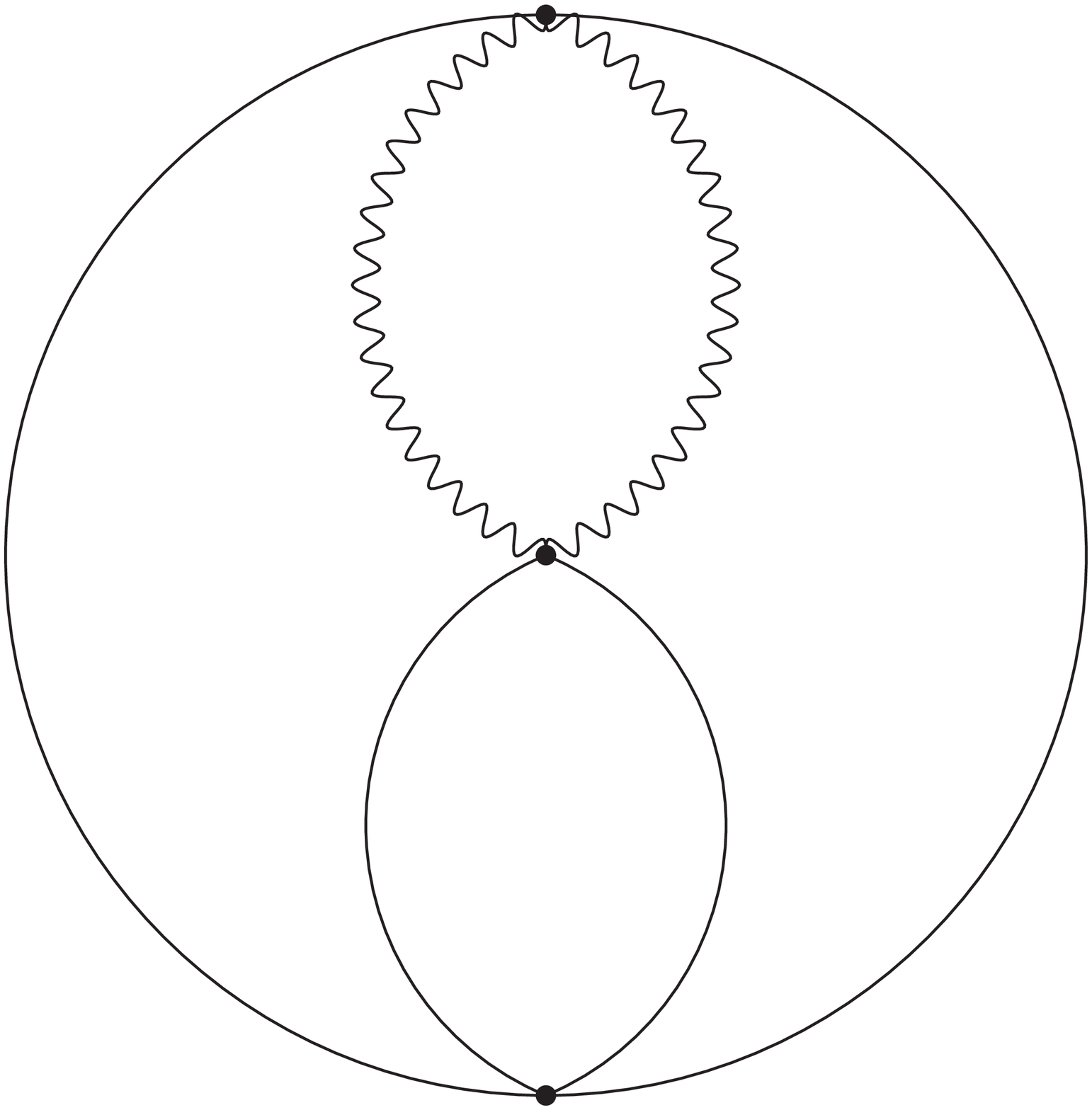}
\vspace{-0.5cm}
\begin{center}
$M_1$
\end{center}
\end{minipage}
\begin{minipage}[t]{2.5cm}
\includegraphics[height=2.5cm,bb=71 107 525 720]{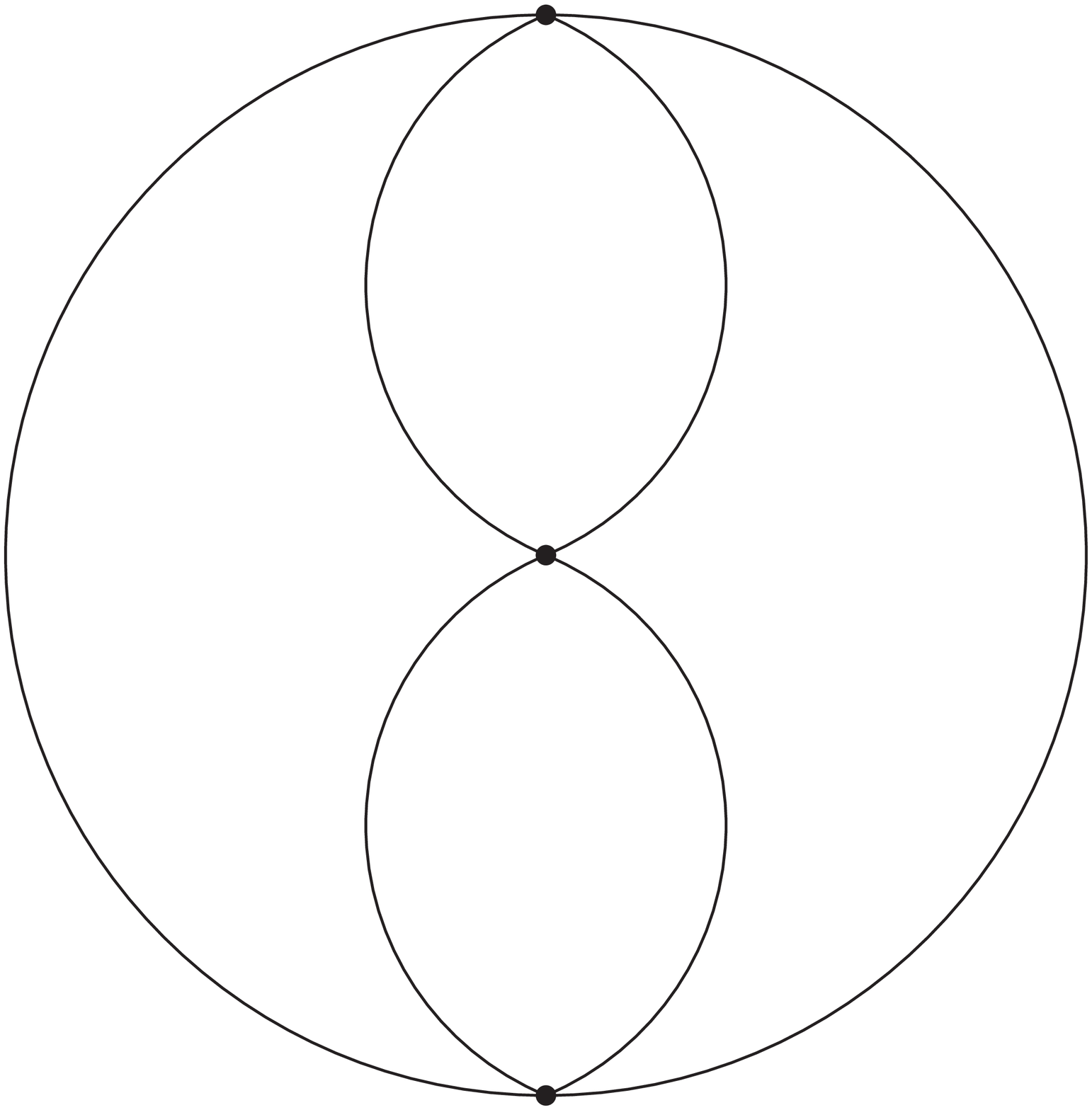}
\vspace{-0.5cm}
\begin{center}
$M_2$
\end{center}
\end{minipage}
\begin{minipage}[t]{2.5cm}
\includegraphics[height=2.5cm,bb=71 107 525 720]{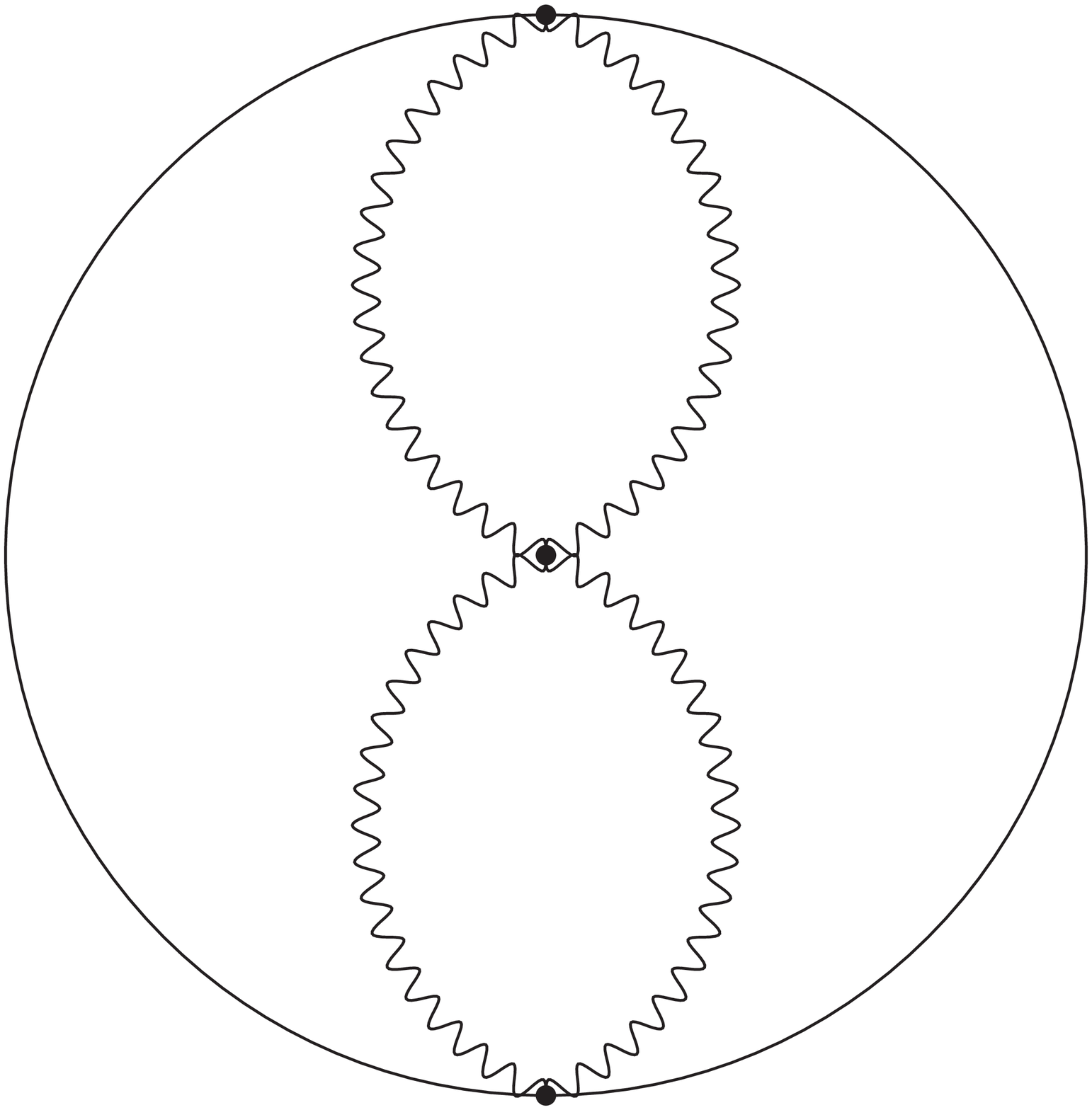}
\vspace{-0.5cm}
\begin{center}
$M_3$
\end{center}
\end{minipage}
\begin{minipage}[t]{2cm}
\includegraphics[height=2.5cm,bb=71 107 525 720]{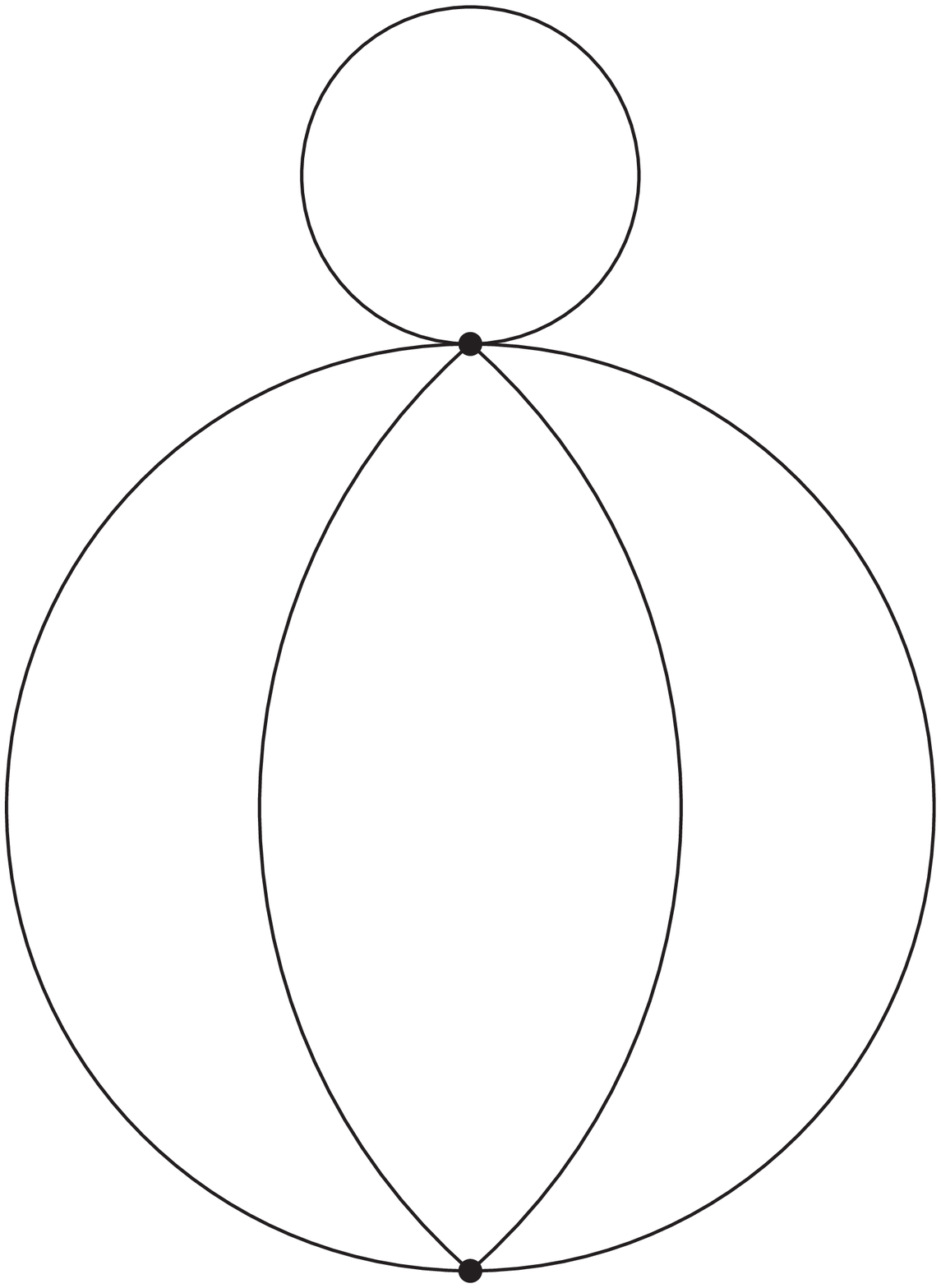}
\vspace{-0.5cm}
\begin{center}
$M_4$
\end{center}
\end{minipage}
\begin{minipage}[t]{2cm}
\includegraphics[height=2.5cm,bb=71 107 525 720]{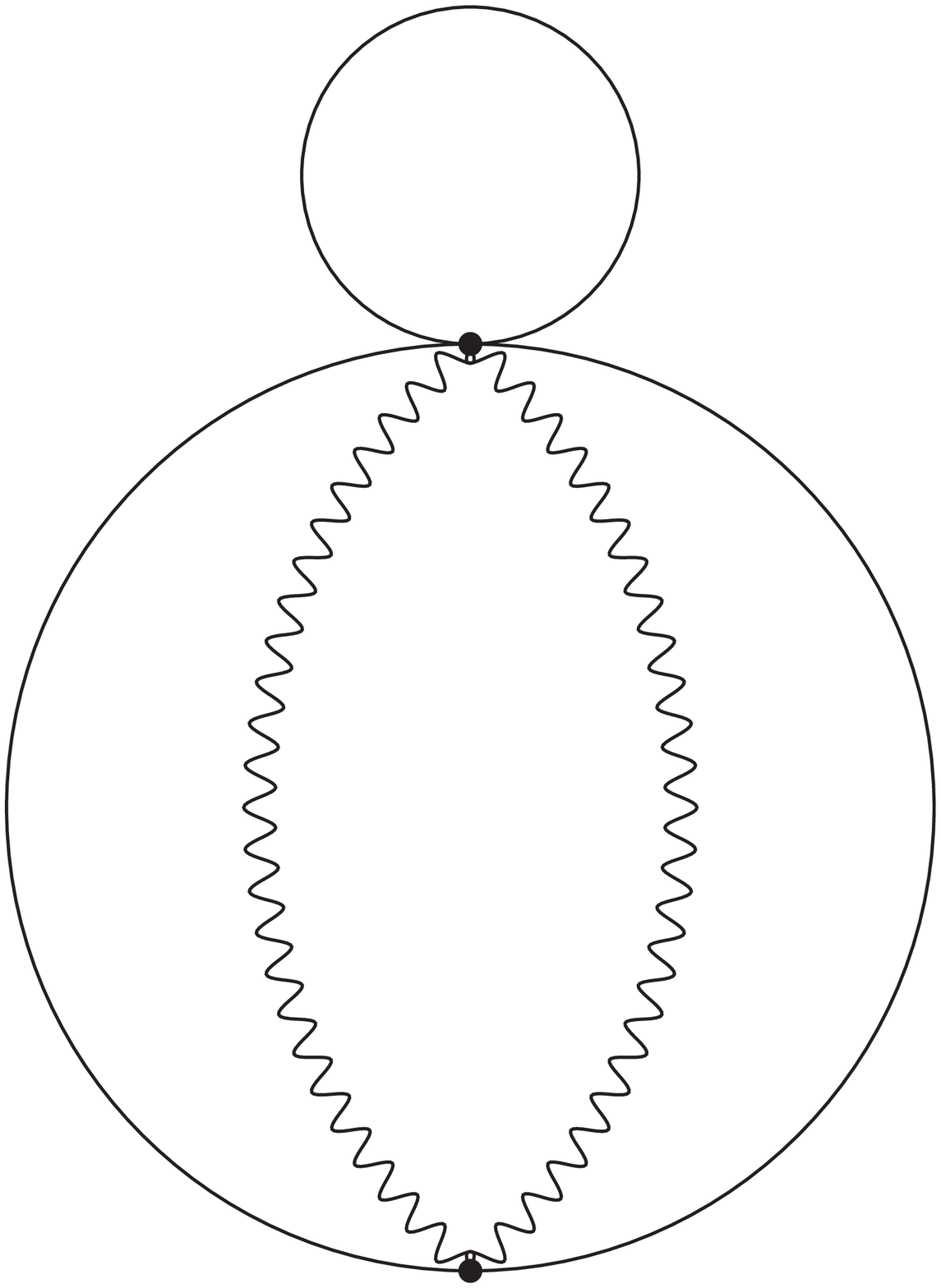}
\vspace{-0.5cm}
\begin{center}
$M_5$
\end{center}
\end{minipage}
\begin{minipage}[t]{2.5cm}
\includegraphics[height=2.5cm,bb=71 107 525 720]{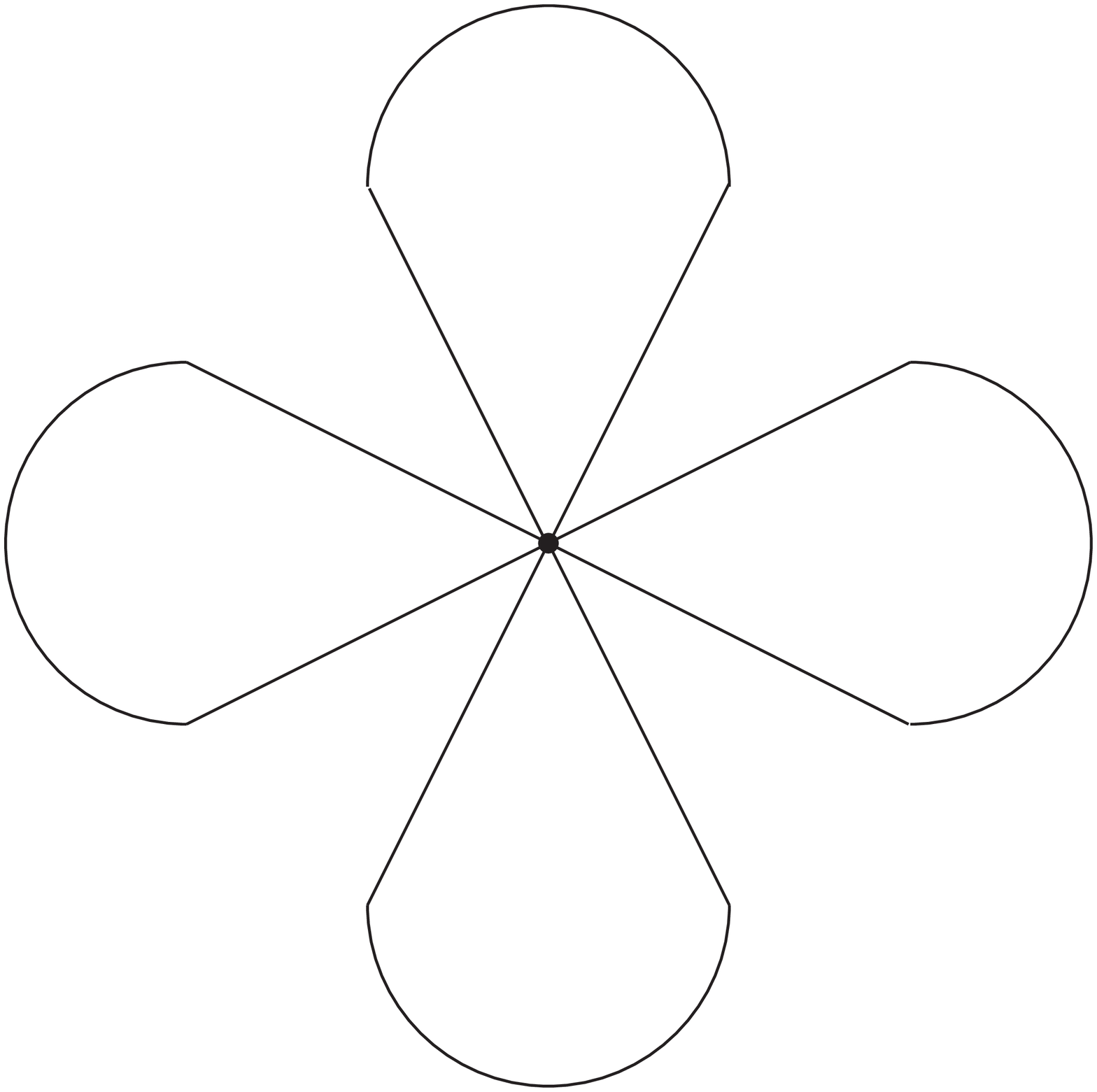}
\vspace{-0.5cm}
\begin{center}
$M_6$
\end{center}
\end{minipage}
\end{center}
\caption{List of independent topologies belonging to the
  $\nf^2$-contribution. The topologies $R_1$-$R_6$ are reducible, whereas
  the topologies $M_1$-$M_6$ are master integrals. The wavy and solid lines
  depict massless and massive propagators respectively.} 
\label{topos}
\end{figure}

Through the expansion in the external momentum $q$ the derivatives
acting on the polarization function generate additional powers of the
denominators of the integrands. The deeper the expansion is, the higher
powers are obtained. Let $M_d$ denote the sum of powers of propagators
$D_{i}$ minus the number of propagators of the generic integral.  Let
$M_p$ denote the total sum of powers of the irreducible scalar products
in the numerator of the integrand. Then one obtains integrands with
$M_d$ up to $8$ and $M_p$ up to $3$, for an expansion of the
$\nf^2$-contribution up to the first physical moment.

For the reduction of these integrals to a set of a few master integrals
the standard method of IBP has been used. The reduction was implemented
following the ideas described in
ref. \cite{Laporta:2001dd,Schroder:2002re,Mastrolia:2000va}. In order to
reduce the polarization function to master integrals for an expansion up
to the first physical moment a system of around 880000
equations has to be generated and solved. For a deeper expansion of the
polarization function one obtains higher values for $M_d$ and $M_p$.
This requires a huger system of IBP-identities in order to obtain a
reduction to master integrals. A lexicographical ordering has been
introduced assigning to each integral a weight describing its
``complexity''. Integrals with increasing powers of the denominator and
increasing number of irreducible scalar products are denoted as
increasingly complicated.  The linear system of equations has been
solved with a program based on FORM3 \cite{Vermaseren:2000nd} which uses
FERMAT \cite{Lewis} for simplifying the rational functions in
the space-time dimension $d$, which arise in this procedure. Complicated
integrals are systematically expressed in terms of simpler ones and then
substituted into the other equations.  Some contributions have been
checked independently with the program SOLVE \cite{SOLVEReference}
from which also some experience concerning the procedure of ordering the
integrals according to their complexity has been gained. Masking of
large integral coefficients is used, a strategy also adopted in the
program AIR~\cite{Anastasiou:2004vj}.

Exploiting the symmetries of the diagrams by reshuffling the powers of
the propagators of a given topology in a unique way strongly reduces the
size of the initial input and, similarly, in the second step the number
of equations which need to be solved.

The solution of the system leads to a set of a around 130000
independent equations.  With the help of these equations the first two
Taylor coefficients of the polarization function can be expressed in
terms of 6 master integrals $M_1$-$M_6$, with denominator powers one and
no irreducible scalar products ($M_{d}=0,\,M_{p}=0$), which are shown in
Fig. \ref{topos}.
\subsection*{Calculation of the master integrals}
The master integrals which belong to the diagrams $M_1-M_6$ are defined
in $d=4-2\*\vep$ space-time dimensions through
\begin{eqnarray}
  M_1&=&{\mu^{16-4\*d}\over N^4} \int \dl1\,\dl2\,\dl3\,\dl4\;
  {1\over D_1\*D_2\*D_4\*D_7\*D_8\*D_{10}},\\
  M_2&=&{\mu^{16-4\*d}\over N^4} \int \dl1\,\dl2\,\dl3\,\dl4\;
  {1\over D_1\*D_2\*D_3\*D_4\*D_9\*D_{10}},\\
  M_3&=&{\mu^{16-4\*d}\over N^4} \int \dl1\,\dl2\,\dl3\,\dl4\;
  {1\over D_4\*D_5\*D_6\*D_7\*D_8\*D_{10}},\\
  M_4&=&{\mu^{16-4\*d}\over N^4} \int \dl1\,\dl2\,\dl3\,\dl4\;
  {1\over D_1\*D_2\*D_3\*D_4\*D_9},\\
  M_5&=&{\mu^{16-4\*d}\over N^4} \int \dl1\,\dl2\,\dl3\,\dl4\;
  {1\over D_2\*D_4\*D_5\*D_7\*D_9},\\
  M_6&=&{\mu^{16-4\*d}\over N^4} \int \dl1\,\dl2\,\dl3\,\dl4\;
  {1\over D_1\*D_2\*D_3\*D_4}
\end{eqnarray}
with
\begin{equation}
  \begin{array}{l@{\quad}l@{\quad}l}
    D_1=\lp_1^2+m^2-i\*\ep, & D_5=\lp_1^2-i\*\ep, &
    D_8=(\lp_1+\lp_2+\lp_3)^2-i\*\ep, \\
    D_2=\lp_2^2+m^2-i\*\ep, & D_6=\lp_2^2-i\*\ep, &
    D_9=(\lp_1+\lp_2+\lp_3)^2+m^2-i\*\ep,\\
    D_3=\lp_3^2+m^2-i\*\ep, & D_7=\lp_3^2-i\*\ep, &
    D_{10}=(\lp_1+\lp_2+\lp_4)^2+m^2-i\*\ep, \\
    D_4=\lp_4^2+m^2-i\*\ep
  \end{array}
\end{equation}
and the normalization factor 
\begin{equation}
  N=\vep\*\mu^{4-d}\int \dl1{1\over D_1^2}
   ={1\over16\*\pi^2}\*\Gamma(1+\vep)\*
                     \left({m^{2}\over4\*\pi\mu^2}\right)^{-\vep}.
\label{NormFac}
\end{equation} 
The factor $\mu$ denotes the renormalization scale. 

Before calculating the master integral $M_1$ we consider at first the 
following combination of integrals with dots, where a dot on a line
denotes an additional power of the associated denominator\\
\vspace{-0.6cm}
\begin{center}
  $m^2$
  \raisebox{-3ex}{\includegraphics[height=1.5cm,bb=71 115 697 720]{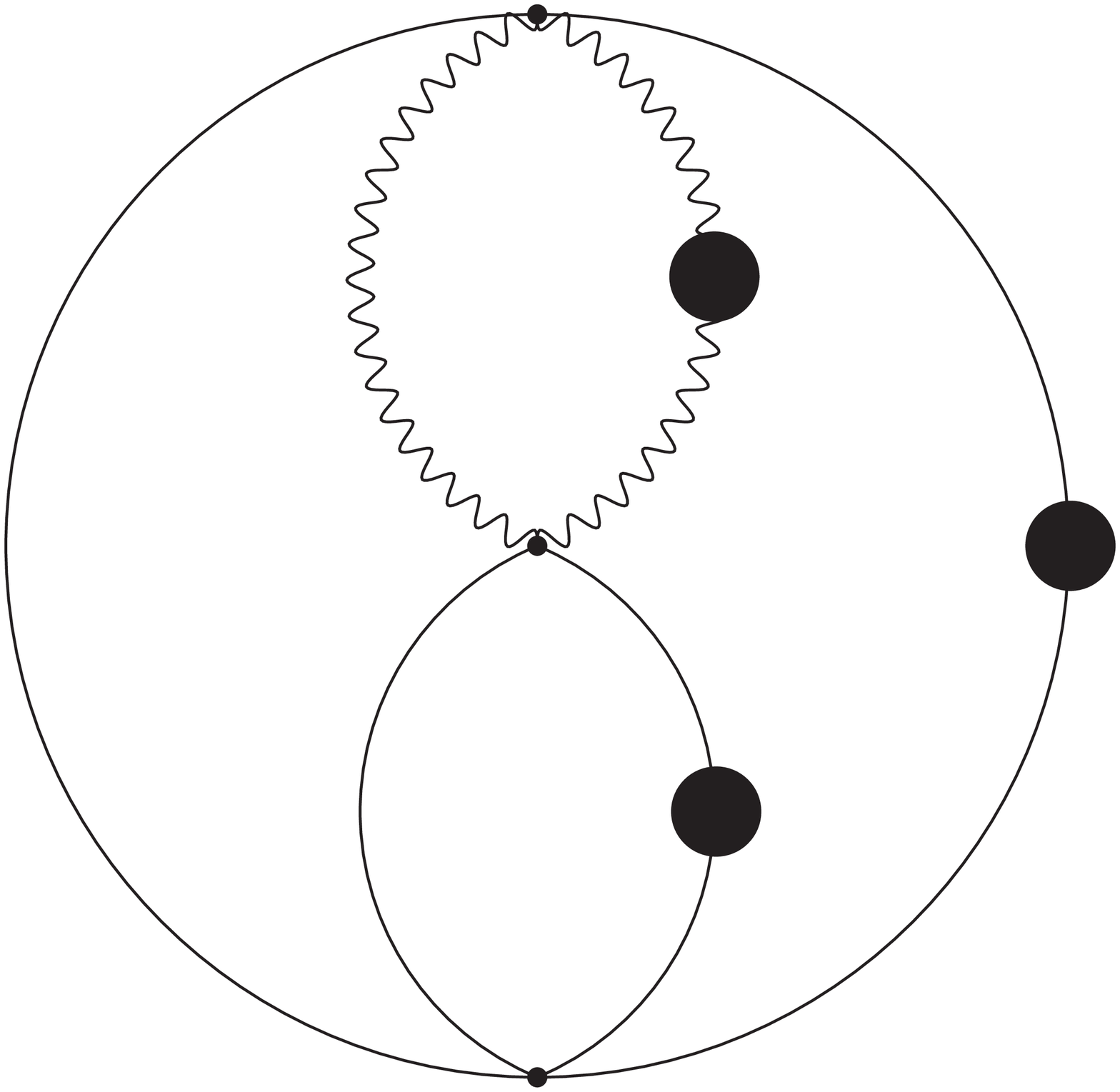}}
  \hspace{0cm}
  $+\;\;{m^2\over\vep}\;$%
  \raisebox{-3ex}{\includegraphics[height=1.5cm,bb=71 114 697 720]{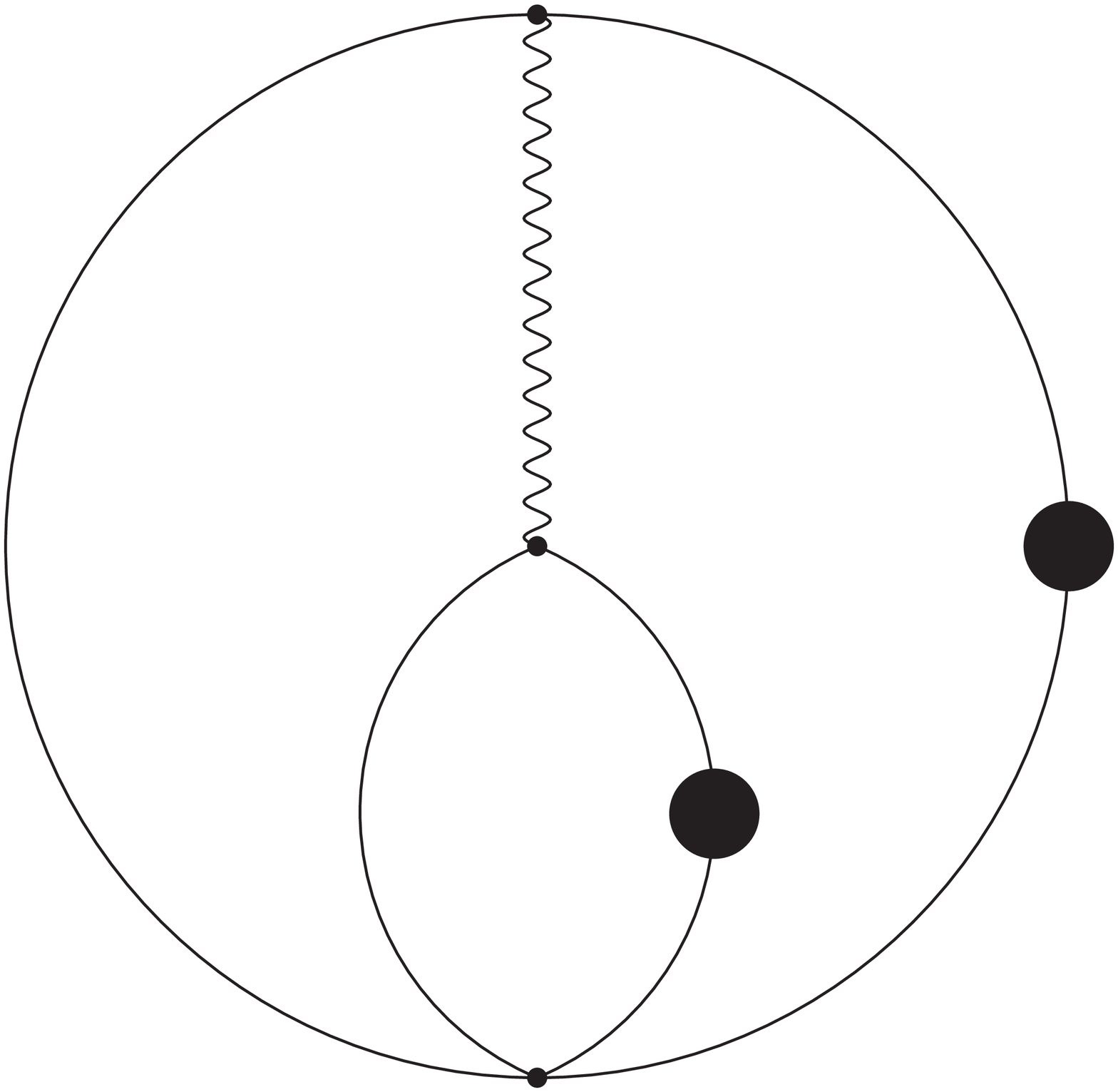}}
  \hspace{0cm}
  $=\;N_{10} + \Order(\vep)$.%
  \vspace{-1.4cm}
  \begin{equation}
    \label{NM1}
  \end{equation}
\end{center}
\mbox{}\\[-1ex] 
The three-loop topology in the left hand side of
eq. (\ref{NM1}) as well as all following three-loop diagrams are
normalized by $N^3$ (of eq.~(\ref{NormFac})).  The combination in
eq. (\ref{NM1}) is finite and can be integrated numerically, with the
result $N_{10}=5.3111546\dots$.
\newpage
\noindent
The relation between the dotted topologies in eq. (\ref{NM1}) and master
integrals can be obtained via IBP. One finds the following relations:
\begin{figure}[!h]
\raisebox{-3ex}{\includegraphics[height=1.5cm,bb=71 115 697 720]{Topo406112With3Dots}}
\hspace{0.4cm}
$={(d-3)^3\*(2\*d-7)\over16\*(d-4)\*m^6}$%
\hspace{0.2ex}
\raisebox{-3ex}{\includegraphics[height=1.5cm,bb=71 107 525 720]{Topo406112}}
\hspace{2ex}
$+\;{(d-3)\*(d-2)\*(3\*d-8)\*(7\*d^2-48\*d+82)\over128\*(d-4)\*(2\*d-7)\*m^8}$
\raisebox{-3ex}{\includegraphics[height=1.5cm,bb=71 107 525 720]{Topo40523}}\hspace{0.5cm},
\vspace{-1.4cm}
\begin{equation}
\label{IBP4M1with3dots}
\end{equation}
\end{figure}
\vspace{-0.5cm}
\begin{figure}[!h]
\raisebox{-3ex}{\includegraphics[height=1.5cm,bb=71 114 697 720]{Topo30503With2Dots}}
\hspace{0.4cm}
$=-{3\*(d-3)\*(3\*d-10)\*(3\*d-8)\over256\*(d-4)\*m^6}$%
\hspace{0.2ex}
\raisebox{-3ex}{\includegraphics[height=1.5cm,bb=71 102 684 720]{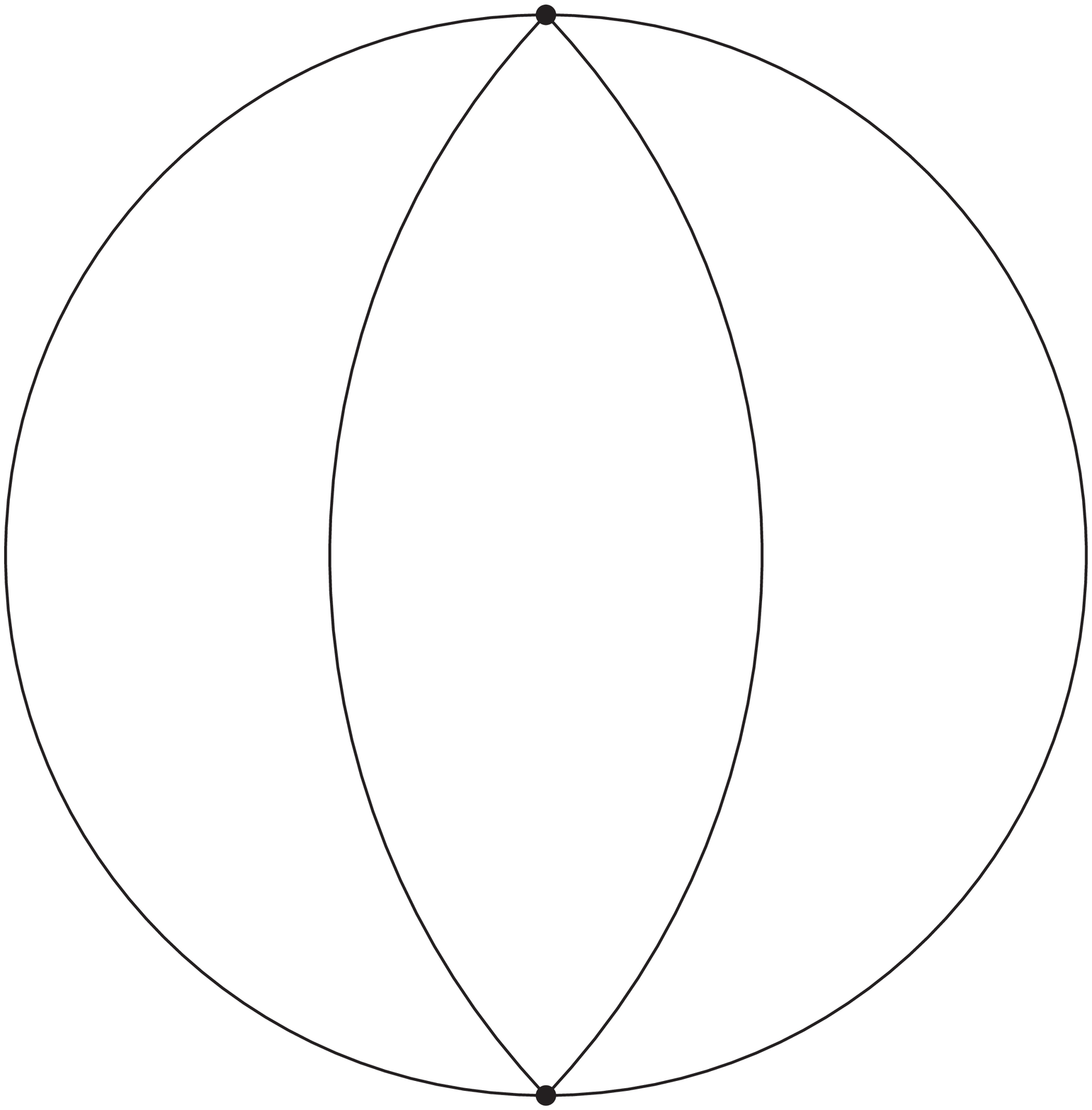}}
$-\;{(d-2)^2\*(11\*d-38)\over128\*(d-4)\*m^8}$
\raisebox{-3ex}{\includegraphics[height=1.5cm,bb=71 178 561 720]{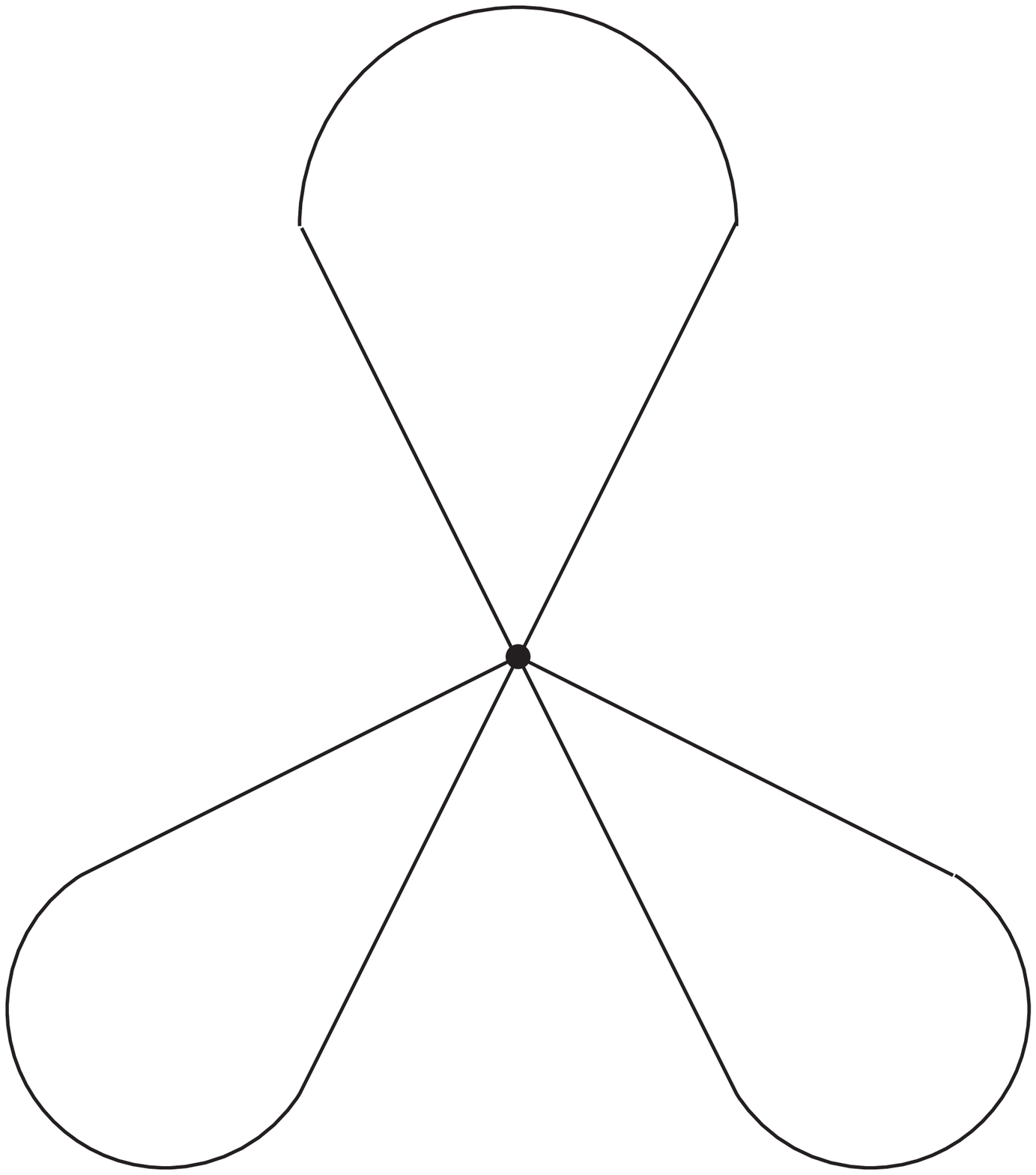}}\hspace{0.5cm}.
\vspace{-1.4cm}
\begin{equation}
\label{3loopwith2dots}
\end{equation}
\end{figure}
\mbox{}\\
The three-loop integrals in the right hand side of
eq. (\ref{3loopwith2dots}) and the factorizable four-loop amplitude in
the right hand side of eq. (\ref{IBP4M1with3dots}) can be calculated 
with MATAD~\cite{Steinhauser:2000ry} or taken from
ref. \cite{Laporta:1997zy}. Inserting eq. (\ref{IBP4M1with3dots}) 
and (\ref{3loopwith2dots}) into eq. (\ref{NM1}) leads~to\footnote{We
  have been informed that the same result has been independently
  obtained in \cite{privatecom}.}
\begin{eqnarray}
  \nonumber
M_1&=&m^4\*\bigg( {2\over3\*\vep^4} + {4\over\vep^3} + {38\over3\*\vep^2} 
   + {4\over3\*\vep}\*\left(11 + 4\*\z3\right) 
   - {2\over15}\*\left(885 + 2\*\pi^4 - 660\*\z3\right) \\
\nonumber
   &-& {4\*\vep\over15}\*(4335 - 1440\*\pl4 - 60\*\Log{2}{4} + 120\*N_{10}
        + 60\*\Log{2}{2}\*\pi^2 \\
   &+& 23\*\pi^4 - 2690\*\z3 - 360\*\z5) + \Order(\vep^2)\bigg),
\end{eqnarray}
with 
\begin{equation}
\hfill\zeta_n=\sum_{k=1}^\infty\,{1\over k^n}
\hspace{0.3cm}\mbox{and}\hspace{0.3cm}
a_4=\sum_{k=1}^{\infty}{1\over2^k k^4}.
\end{equation}
The same procedure has been applied for calculating the master integral
$M_2$. The topology $M_2$ with 3 symmetrical distributed dots is finite\\
\begin{center}
\vspace{-0.6cm}
$m^2\*$
  \raisebox{-3ex}{\includegraphics[height=1.5cm,bb=71 114 697 720]{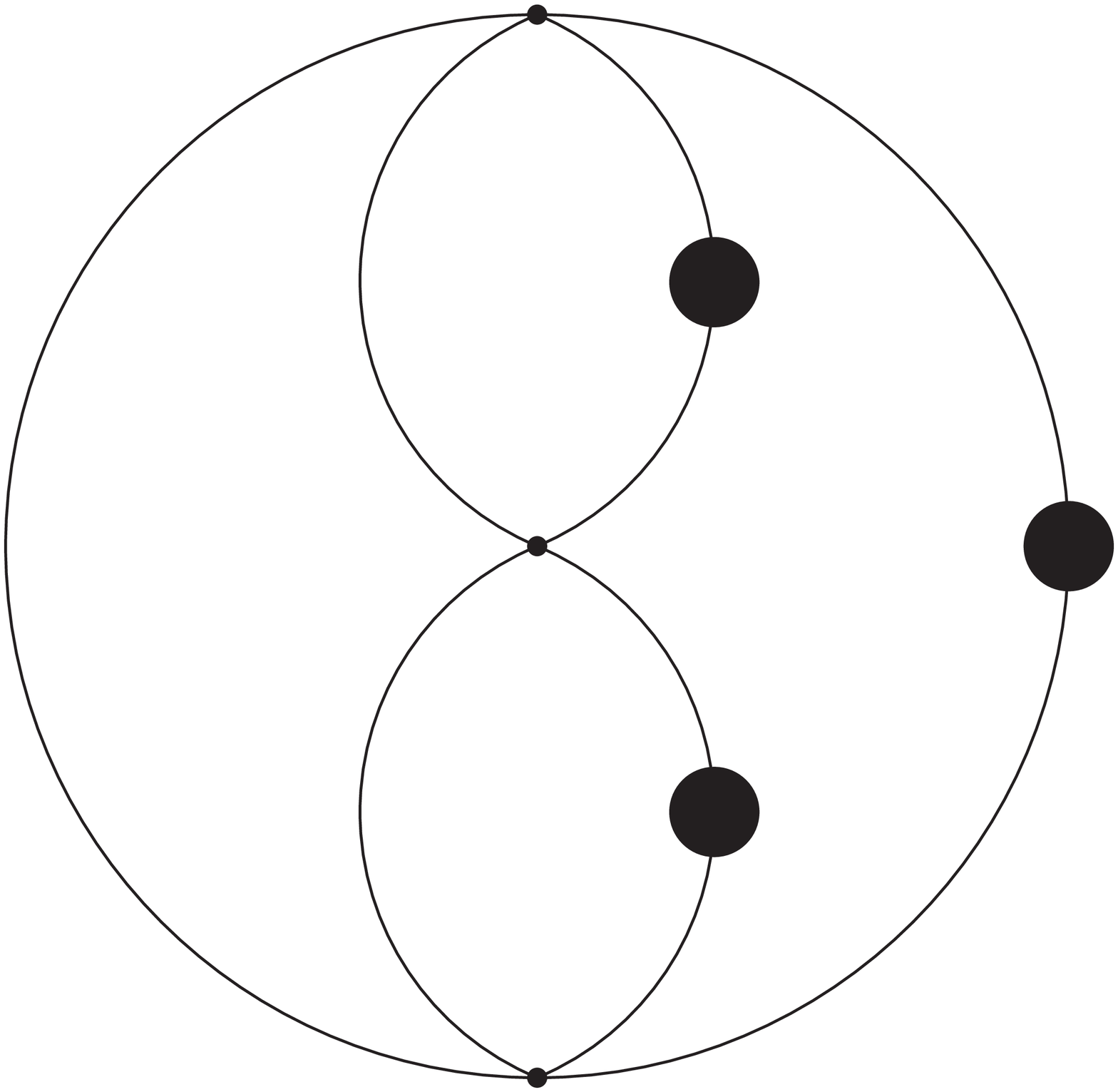}}
\hspace{0cm}
$=N_{20} + \Order(\vep)$
\end{center}
\vspace{-1cm}
\begin{equation}
\label{NM2}
\end{equation}\\[-2ex]
and can be integrated numerically. The calculation yields
$N_{20}=5.40925606\dots$.
The master integral $M_2$ can be calculated using the IBP identity \\[1ex]
\raisebox{-3ex}{\includegraphics[height=1.5cm,bb=71 114 697 720]{Topo406111With3Dots}}
\hspace{0cm}
$=-\;{(d-4)\*(d-3)^3\*(2\*d-7)\over6\*(3\*d-13)\*(3\*d-11)\*m^6}$
  \raisebox{-3ex}{\includegraphics[height=1.5cm,bb=71 107 525
      720]{Topo406111}}
\hspace{0.3cm}
$-\;{(d-3)\*(d-2)\*(3\*d-8)\*(13\*d^2-99\*d+182)\over
   512\*(3\*d-13)\*(3\*d-11)\*m^8}$
  \raisebox{-3ex}{\includegraphics[height=1.5cm,bb=71 107 525
      720]{Topo40524}}
\hspace*{1.8cm}
$+\;{(d-2)^4\*(d-1)\over256\*(3\*d-13)\*(3\*d-11)\*m^{10}} $
  \raisebox{-3ex}{\includegraphics[height=1.5cm,bb=71 107 525 720]{Topo40401}}\hspace{0.5cm},
\vspace{-1cm}
\begin{equation}
\label{IBP4MI2}
\end{equation}
inserting eq. (\ref{IBP4MI2}) into eq. (\ref{NM2}) and solving with respect
to $M_2$ results in: 
\begin{eqnarray}
\nonumber
M_2&=&m^4\*\bigg( {3\over2\*\vep^4} + {19\over2\*\vep^3} 
                + {67\over2\*\vep^2} 
                + {1\over2\*\vep}\*\left(127-6\*N_{20}+21\*\z3\right)
                + \Order(\vep^0)\bigg).
\end{eqnarray}
This result is in agreement with eq. (4) in
ref. \cite{Laporta:2002pg}.\\
The calculation of the master integral $M_3$ is easy, 
\begin{eqnarray}
\nonumber
M_3&=&m^4\*\bigg({1\over6\*\vep^4} + {5\over6\*\vep^3} + {3\over2\*\vep^2} 
     - {1\over6\*\vep}\*(39-38\*\z3) 
     - {1\over60}\*(4710+23\*\pi^4\\
    &-&1900\*\z3) 
     - {\vep\over12}\*\left(5934 +23\*\pi^4 - 684\*\z3 - 2580\*\z5\right)
     +\Order(\vep^2)\bigg).
\end{eqnarray}
For completeness we also give the results for the factorized
master integrals
\begin{eqnarray*}
\nonumber
M_4&=&m^6\*\bigg(-{2\over\vep^4}-{29\over3\*\vep^3}-{163\over6\*\vep^2}
            -{601\over12\*\vep}
            - {1\over24}\*\left(635+896\*\z3\right)
    + {\vep\over720}\*(204705\\
    &-&184320\*\pl4 
     -7680\*\Log{2}{4} + 7680\*\Log{2}{2}\*\pi^2+2176\*\pi^4-228480\*\z3)
    +\Order(\vep^2)\bigg),\\
\nonumber
M_5&=&m^6\*\bigg(-{1\over3\*\vep^4}-{3\over2\*\vep^3}-{43\over12\*\vep^2} 
                 -{1\over24\*\vep}\*\left(81+64\*\z3\right)
     + {1\over240}\*(3985+32\*\pi^4\\
     &-&2880\*\z3)
     +{\vep\over480}\*\left(60435+288\*\pi^4
                 -13760\*\z3-23040\*\z5\right)+\Order(\vep^2)\bigg),\\
M_6&=&m^8\*\left( {1\over\vep^4}+{4\over\vep^3}+{10\over\vep^2}
           +{20\over\vep}+35+56\*\vep+\Order(\vep^2)\right).
\end{eqnarray*}

\subsection*{Result for the {\boldmath{$\Order(\als^3 n_f^2)$}} contribution}
Inserting the above master integrals into the reduced
$n_{f}^2$-contribution and performing renormalization of the strong
coupling constant $\als$, the external current and the mass
$m=\overline{m}(\mu^2)$ in the $\overline{\mbox{MS}}$-scheme, leads to
the following result for the first two moments of the four-loop
$\Order(\als^3 n_f^2)$ contribution of the heavy quark vacuum
polarization function:
\begin{eqnarray}
\nonumber
\hat{C}^{(3)}_{0}&=&\nl\*\nh\*
           \Bigg({7043\over11664} - {127\over108}\*\z3 + {1\over6}\*N_{10}
           + \left({37\over324} - {7\over24}\*\z3\right)\*\lmusdms         
           + {2\over27}\*\lmusdms^2 - {2\over27}\*\lmusdms^3 
	 \Bigg)\\ \nonumber
        &+&\nh^2\*
           \Bigg({610843\over816480} + {1439\over540}\*\z3 
           - {157\over210}\*N_{20} 
           + \left({113\over324} - {7\over24}\*\z3\right)\*\lmusdms 
           + {1\over27}\*\lmusdms^2 - {1\over27}\*\lmusdms^3
           \Bigg)\\ \nonumber
        &+&\nl^2\*
           \Bigg({17897\over23328} - {31\over54}\*\z3
          - {19\over81}\*\lmusdms +{1\over27}\*\lmusdms^2 
          - {1\over27}\*\lmusdms^3
           \Bigg)\\ 
\mbox{and}&&\\ \nonumber
\nonumber
\hat{C}^{(3)}_{1}&=&\nl\*\nh\*
         \Bigg({262877\over262440} - {38909\over19440}\*\z3 
          + {29\over81}\*N_{10} 
        \\&&\qquad \nonumber
	  + \left({3779\over21870} - {203\over324}\*\z3\right)\*\lmusdms
	  + {472\over3645}\*\lmusdms^2 - {16\over135}\*\lmusdms^3  
           \Bigg)\\ \nonumber
        &+&\nh^2\*
         \Bigg({163868\over98415} + {13657\over2430}\*\z3
             - {5648\over3645}\*N_{20} 
        \\&&\qquad \nonumber
         + \left({14483\over21870} - {203\over324}\*\z3\right)\*\lmusdms 
         + {236\over3645}\*\lmusdms^2 - {8\over135}\*\lmusdms^3 
         \Bigg)\\ 
        &+&\nl^2\*
         \Bigg({42173\over32805} - {112\over135}\*\z3
         - {1784\over3645}\*\lmusdms + {236\over3645}\*\lmusdms^2 
         - {8\over135}\*\lmusdms^3 \Bigg)
\end{eqnarray}
with $\lmusdms=\log\left({\mu^2\over m^2}\right)$.\\
The $\nl^2$-contribution has been checked independently by taking into
account the corresponding two-loop case, in which the gluon propagator
has been replaced by a gluon propagator containing a renormalon chain
with two massless fermion one-loop insertions. The computation has been
performed in a general $\xi$-gauge and it has been checked that the
dependence on the gauge parameter vanishes.  Furthermore it has been
checked that both coefficient functions $C_0$ and $C_1$ meet the
standard renormalization group equation.  Numerically one finds for the
coefficient $C_{0}$ and $C_{1}$ at $\mu=m$:
\begin{eqnarray}
C_{0}&=&
       a_{s}\*1.4444
      +a_{s}^2\*\left(1.5863 + 0.1387\*\nh + 0.3714\*\nl\right)
      \\\nonumber
      &+&a_{s}^3\*
      \left(0.0252\*\nh\*\nl + 0.0257\*\nl^2 - 0.0309\*\nh^2\right),\\
C_{1}&=&1.0667
      +a_{s}\*2.5547
      +a_{s}^2\*\left(0.2461 + 0.2637\*\nh + 0.6623\*\nl\right)
      \\\nonumber
      &+&a_{s}^3\*
      \left( 0.1658\*\nh\*\nl + 0.0961\*\nl^2 + 0.0130\*\nh^2\right).
\end{eqnarray}
\section{Summary and Conclusion}
\label{DiscussConclude}
Using the IBP method and Laporta's algorithm, we have evaluated a gauge
invariant subset of the four-loop tadpole amplitudes contributing to
derivatives of the vacuum polarization at $q^2=0$.  All loop integrals
have been mapped on a minimal set of independent topologies. Then an
elaborate automated procedure has been developed and applied, which
identifies equivalent amplitudes, factorizable contributions, discards
massless tadpoles, and performs symmetrization.  Solving a system of
nearly one million linear equations, all amplitudes can be expressed
through six master integrals. These have been evaluated analytically or
numerically to high precision. The present work can be seen as a first
step towards the evaluation of the full set of four-loop amplitude
contributions to the vacuum polarization function.\\

\parindent0cm
{\bf Acknowledgments:}\\
We would like to thank Mikhail Tentyukov for generous support and a lot
of advice in our dealing with FORM and FERMAT. Furthermore we thank
Matthias Steinhauser for careful reading the manuscript and
advice. We are grateful to Michael Faisst for discussions.
C.S. would like to thank the Graduiertenkolleg {\em ``Hochenergiephysik
und Teilchenastrophysik''} for financial support. The work of P.M. was
partially supported by the Marie Curie Training Site {\em ``Particle
Physics at Present and Future Colliders''}.  This work was supported by
the SFB/TR 9 (Computational Particle Physics).
\begin{appendix}

\end{appendix}

\end{document}